\documentclass[aip,jmp,amsmath,amssymb,reprint]{revtex4-1}

\usepackage{graphicx}
\usepackage{dcolumn}
\usepackage{bm}
\usepackage{siunitx}

\usepackage{import}
\usepackage{hyperref}

\usepackage{array}

\graphicspath{ {Figures/} }

\begin{document}

\def\mca#1{\multicolumn{1}{c}{#1}}
\def\mcb#1{\multicolumn{1}{c|}{#1}}
\renewcommand{\arraystretch}{2.25}


\title[Studies of MCP-PMTs in the miniTimeCube neutrino detector]{Studies of MCP-PMTs in the miniTimeCube neutrino detector}

\author{V.A.~Li}\email[Electronic mail:\;]{vli2@hawaii.edu}
\affiliation{Department of Physics and Astronomy, University of Hawai`i at M\=anoa}
\author{J.~Koblanski}\email[Electronic mail:\;]{johnk2@hawaii.edu}
\affiliation{Department of Physics and Astronomy, University of Hawai`i at M\=anoa}
\author{R.~Dorrill}
\affiliation{Department of Physics and Astronomy, University of Hawai`i at M\=anoa}
\author{M.J.~Duvall}
\affiliation{Department of Physics and Astronomy, University of Hawai`i at M\=anoa}
\author{K.~Engel}
\affiliation{Department of Physics, University of Maryland}
\author{G.R.~Jocher}
\affiliation{Ultralytics LLC}
\author{J.G.~Learned}
\affiliation{Department of Physics and Astronomy, University of Hawai`i at M\=anoa}
\author{S.~Matsuno}
\affiliation{Department of Physics and Astronomy, University of Hawai`i at M\=anoa}
\author{W.F.~McDonough}
\affiliation{Department of Geology, University of Maryland}
\author{H.P.~Mumm}
\affiliation{National Institute of Standards and Technology}
\author{S.~Negrashov}
\affiliation{Department of Physics and Astronomy, University of Hawai`i at M\=anoa}
\affiliation{Department of Information and Computer Sciences, University of Hawai`i at M\=anoa}
\author{K.~Nishimura}
\affiliation{Department of Physics and Astronomy, University of Hawai`i at M\=anoa}
\author{M.~Rosen}
\affiliation{Department of Physics and Astronomy, University of Hawai`i at M\=anoa}
\author{M.~Sakai}
\affiliation{Department of Physics and Astronomy, University of Hawai`i at M\=anoa}
\author{S.M.~Usman}
\affiliation{Department of Geography and Geoinformation Science, George Mason University}
\author{G.S.~Varner}
\affiliation{Department of Physics and Astronomy, University of Hawai`i at M\=anoa}
\author{S.A.~Wipperfurth}
\affiliation{Department of Geology, University of Maryland}

\date{\today}

\begin{abstract}
\textbf{Abstract\newline}
This report highlights two different types of cross-talk in the photodetectors of the miniTimeCube neutrino experiment.
The miniTimeCube detector has 24 $8 \times 8$-anode Photonis MCP-PMT Planacon XP85012, totalling 1536 individual pixels viewing the 2-liter cube of plastic  scintillator.
\end{abstract}


\text{Published in \href{https://doi.org/10.1063/1.5043308}{AIP Advances 8, 095003 (2018)}.}
                              
\maketitle



\section{\label{sec:level1}Introduction: miniTimeCube experiment and MCP-PMT readout electronics}

\begin{figure}[h]
    \centering
    \includegraphics[width=0.5\textwidth]{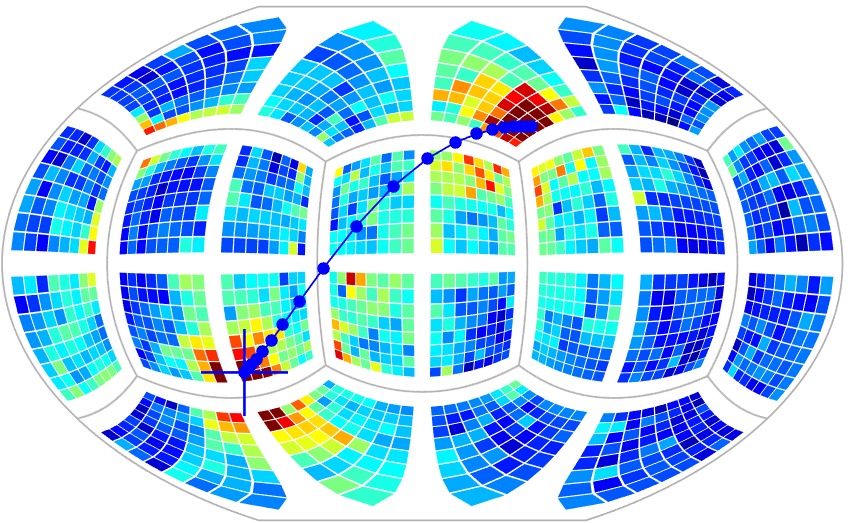}
    \caption{Charge distribution and reconstructed muon track in a typical muon event, observed in the 1536-pixel mTC. }
    \label{fig:mTC}
\end{figure}

The miniTimeCube (mTC) is a compact mobile anti-neutrino prototype detector previously deployed at the NIST nuclear reactor~\cite{Li:2016yey,Li:2018dwm}. 
The mTC detector is a 2-liter cube of plastic scintillator EJ-254 doped with 1\% natural boron.
Four micro-channel-plate photomultiplier tubes (MCP-PMTs) cover each face of the cube, 
and each MCP-PMT has 64 individual anodes.
The MCP-PMTs are coupled to the scintillator with an optical grease and are secured with plastic clamps to a plastic frame surrounding the scintillator.
There are 24 MCP-PMTs in total, therefore yielding 1536 individual ``pixels'' viewing the scintillator. 
To read out the signals from the MCP-PMTs, modules of compact sub-100-ps fast-timing  electronics are mounted on the back of the MCP-PMTs. Fig.~\ref{fig:mTC} shows how a typical muon event presents in the real mTC data due to this MCP-PMT setup.

During the operation of the detector, it was observed that there 
were two types of cross-talk, both between neighboring tubes and neighboring pixels on individual tubes, which we call cross-talk 1.0 and cross-talk 2.0 respectively. 


\section{Cross-talk 1.0}
\begin{figure}[h]
    \centering
    \includegraphics[width=0.48\textwidth]{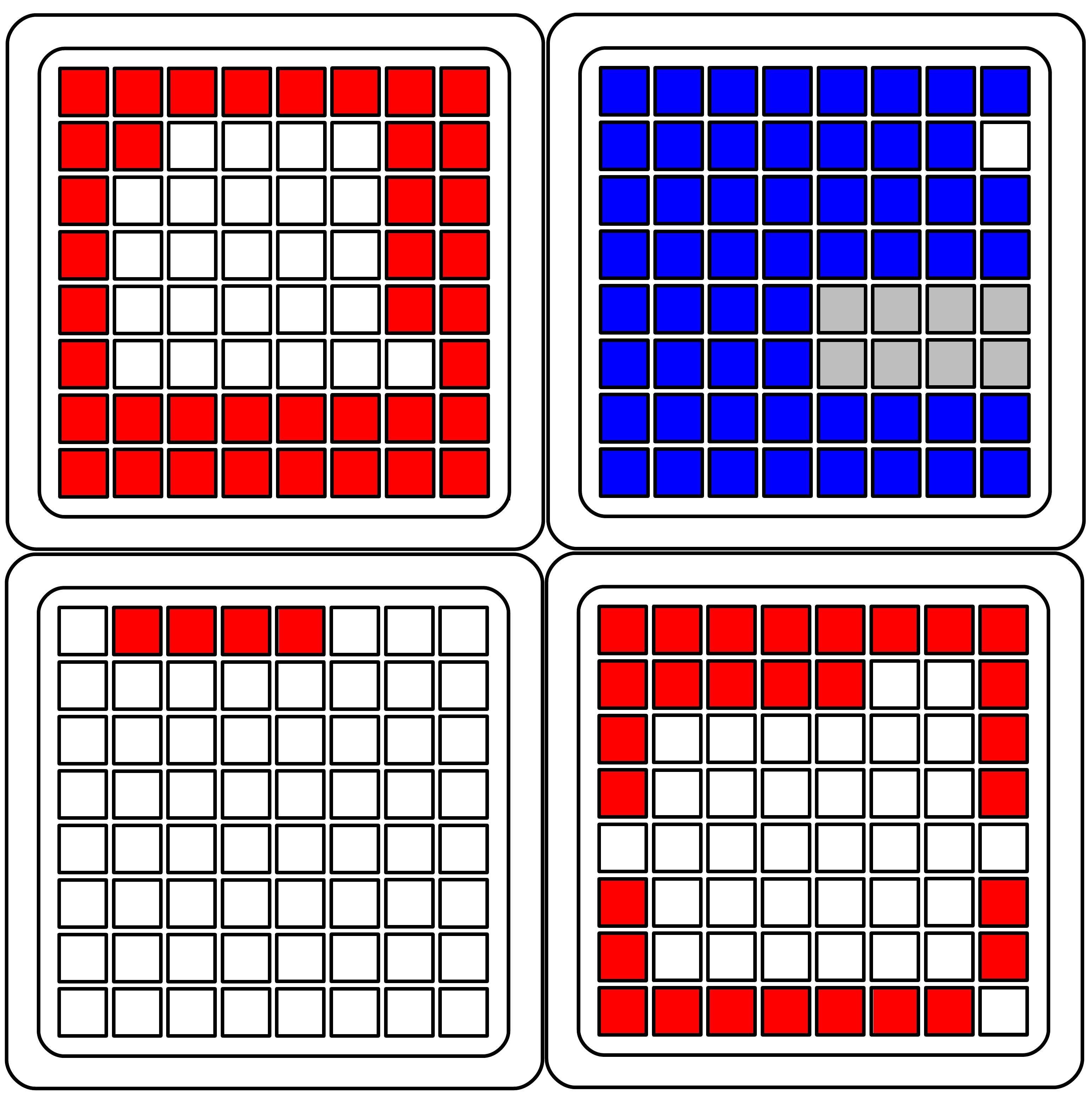}
    \caption{An example of cross-talk 1.0.
    Four MCP-PMTs are mounted on one face. Only the top right corner MCP-PMT is powered; all other MCP-PMTS in the entire mTC are unpowered.
    In this event the anodes (pixels), colored in red and blue, have waveforms, while the white squares remain  untriggered. Red squares are pixels with waveforms on the {\it unpowered} tubes; blue --- pixels with waveforms on the {\it powered} tube. The 8 gray channels are non-responsive due to a dead ASIC. }
    \label{fig:MCPdia1}
\end{figure}

\begin{figure}[h]
    \centering
    \includegraphics[width=0.48\textwidth]{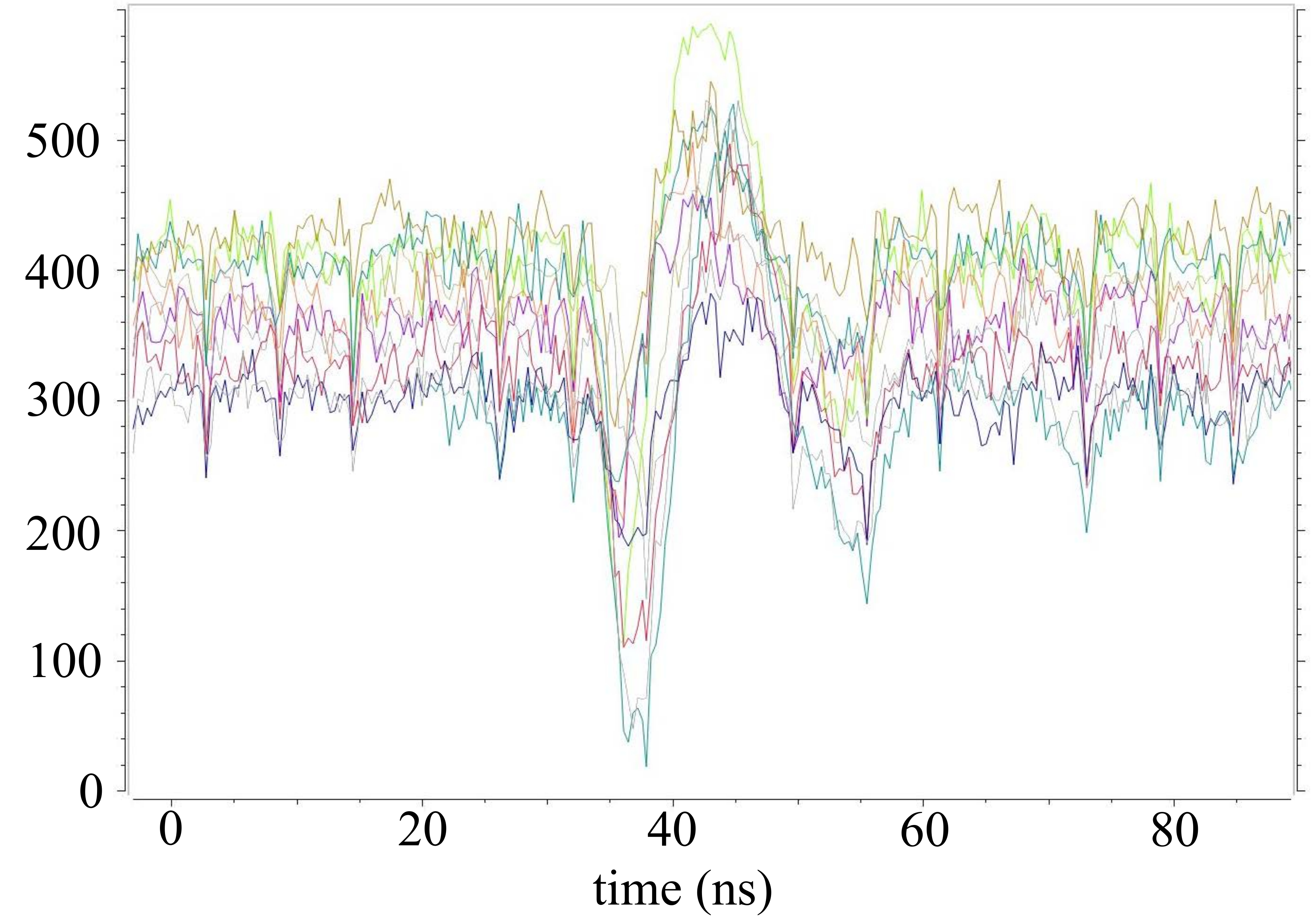}
    \caption{Waveforms on the edge pixels of the unpowered MCP-PMTs, neighboring one powered MCP-PMT. Y-axis is in ADC counts.}
    \label{fig:phantom1}
\end{figure}

This type of cross-talk was accidentally discovered during gain-calibration tests. Only one tube out of 24 was powered at a time. 
Fig.~\ref{fig:MCPdia1} is an example of one such event. 
The edge pixels of the {\it unpowered} MCP-PMTs, neighboring the one {\it powered} MCP-PMT, had {\it phantom} waveforms.
The phantom pulses were a few hundred counts high, shown in Fig.~\ref{fig:phantom1}, and had a distinct negative-going pulse preceding a positive-going peak, 
which led us to believe there might have been 
a capacitive coupling between the tubes. 
Large pulses caused by muons don't have such pre-pulsing features.
Moreover, as it turned out, we are the only customer of Photonis who uses their MCP-PMTs in such a densely-packed configuration with four tubes next to each other. 
We removed the readout modules from the back of the MCP-PMTs, and injected different RF pulses into individual pins, as well as into an MCP output. 
Our tests confirmed that the effect is not due to the electronics or disturbances in the HV system~\cite{Li:2018dwm}. 
This effect is new and hasn't been discussed in the literature.
 
It also became apparent that this effect affects the timing performance of the detector, since in reality, when everything is powered up, the waveforms on the edge pixels are superpositions of the phantom and real pulses. The leading edge becomes quite distorted, especially in the case of dim signals.
The detector would also be triggered on those pulses as they were quite large. 

\begin{figure}[h]
    \centering
    \includegraphics[width=0.48\textwidth]{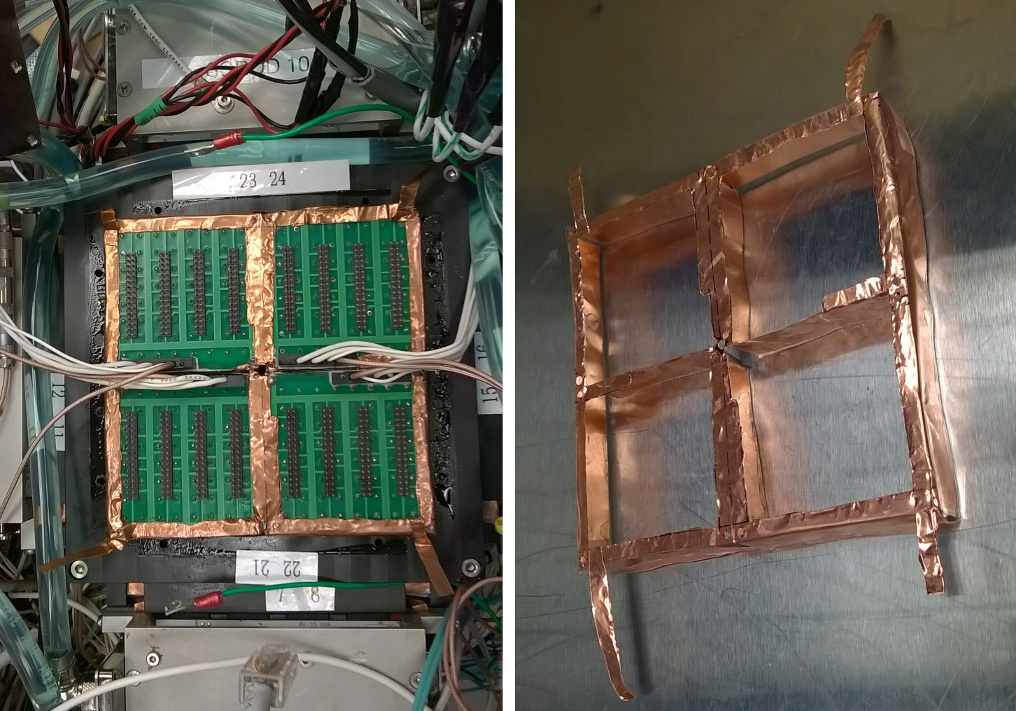}
    \caption{Copper frame. Copper strips at the corners will be bent and grounded to the electronics card cages.}
    \label{fig:copper_frame_combined}
\end{figure}

Many ideas, which were intended to rectify this effect, were explored, and a complete solution to the  problem was achieved as follows. A special frame made of .004$''$-thick 99.97\%-pure copper was designed and manufactured, shown in Fig.~\ref{fig:copper_frame_combined}.
The frame goes all the way around the perimeters of all four MCP-PMTs sharing one face. 
Another key feature of the frame is that it touches four grounding pins at the four corners of each MCP-PMT and grounds to the electronics card cages. 
The grounding cables of the MCP-PMTs were also removed, as they contributed to the cross-talk on a smaller level.

%


\section{Cross-talk 2.0}

A completely different phenomenon was observed while performing single-PE tests.
Although at first it resembles a cross-talk between neighboring pixels within a single MCP-PMT,  the nature of this phenomenon is not a cross-talk; however, we refer to it as cross-talk 2.0, for the sake of brevity.

\begin{figure}[h]
\centering
\includegraphics[width=.48\textwidth]{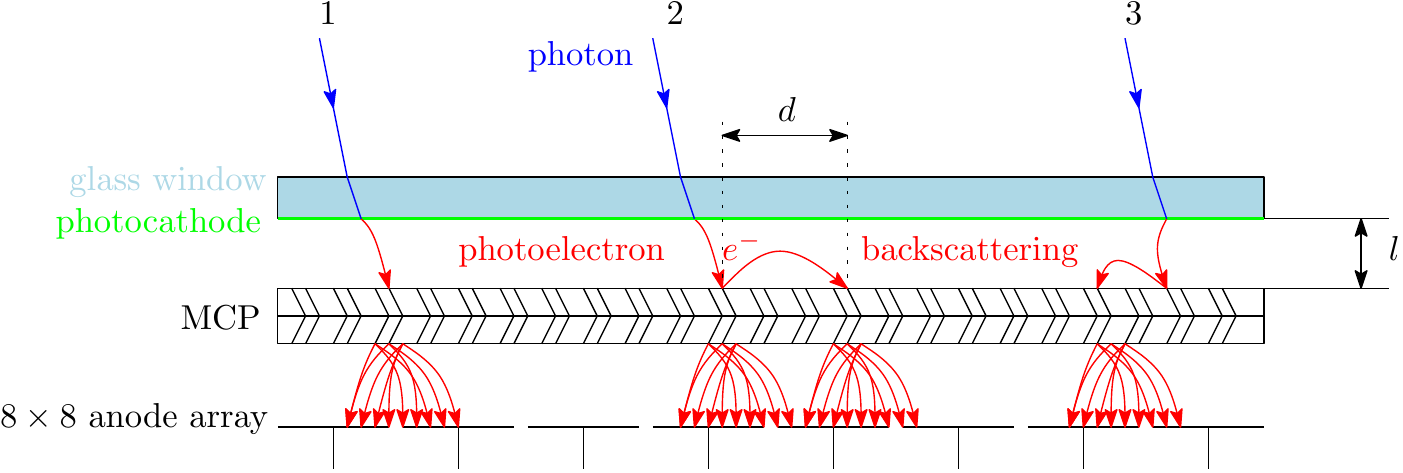}
\caption{MCP-PMT diagram showing three different charge sharing scenarios due to the spread of the electron shower (1) and  inelastic (2) and elastic (3) $e^-$ backscattering.\label{fig:charge_sharing}}

\end{figure}

At low light output in the laser data, we would often see a single pixel light up accompanied by its neighbors rather than by itself. 
After performing more tests and researching the literature~\cite{INAMI2008247,Vavra:2010zz},
we realized that it is a charge-sharing effect within the multi-anode structure of MCP-PMTs.
Fig.~\ref{fig:charge_sharing} provides an illustration of the charge-sharing effect caused by the spread of the electron shower after the MCP and elastic (or inelastic) photoelectron back-scattering. 
It is worth noting here that a similar problem was solved by Hamamatsu~\cite{INAMI2008247} by segmenting the MCP electrode.

The charge sharing can also introduce a delay in the waveform.
The maximum delay and distance can be estimated to be~\cite{Korpar11}
\begin{equation}
 t_{max} \approx 2 t_0, \qquad \qquad d \approx 2 l
\end{equation}
where $t_0 = l\sqrt{2m_e/(e V)}$ is the approximate time of flight for a photoelectron from the photocathode to the MCP. Both elastic (longer delay) and inelastic (shorter delay) scattering off the MCP are possible.
If the distance between photocathode and MCP is on the order of a few mm, then we get the next pixel within the range for backscattered photoelectrons to reach a neighboring pixel, thus creating cross-talk and delay on the order of nanoseconds.

\begin{figure}[h]
\centering
\includegraphics[width=.48\textwidth]{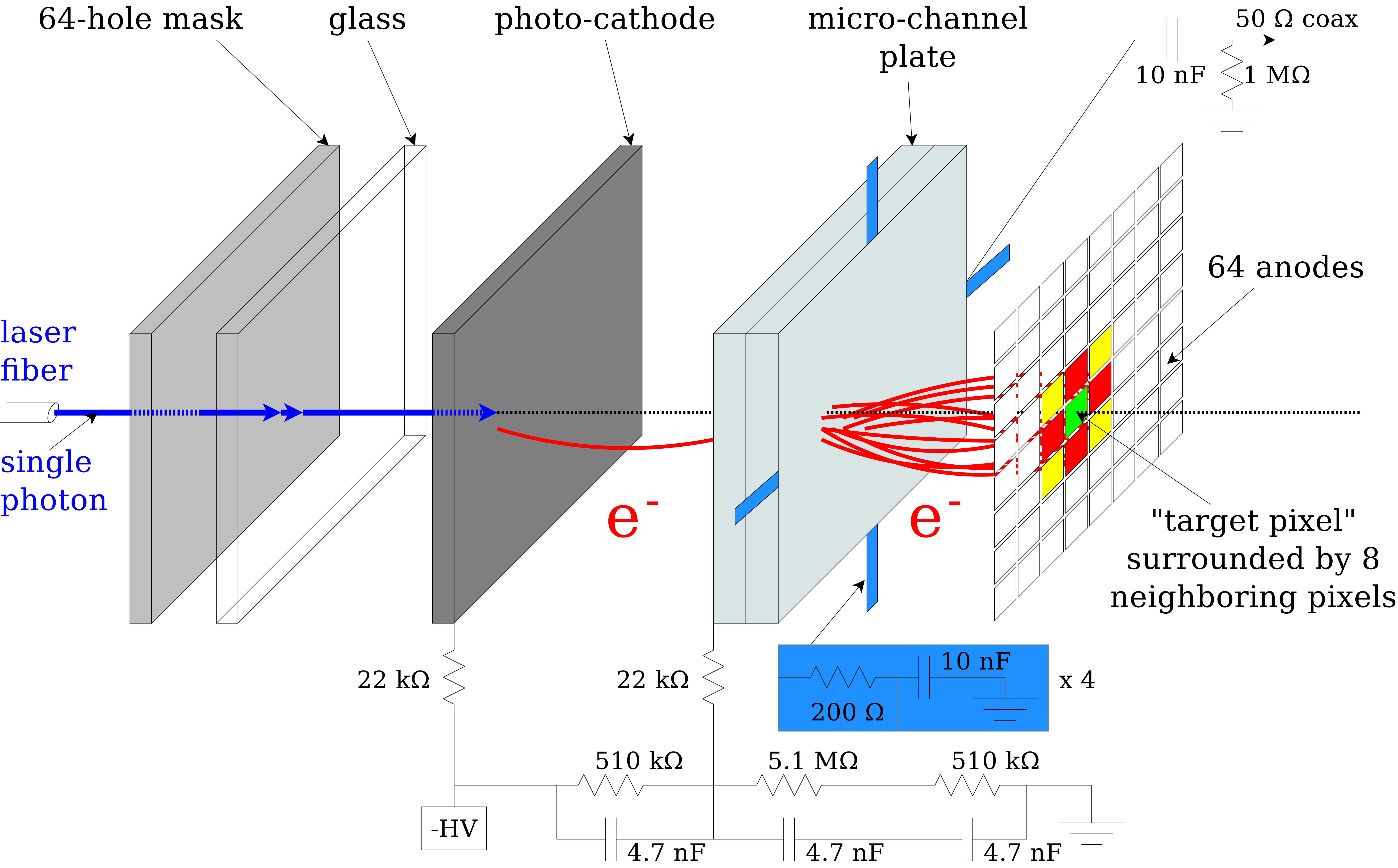}
\caption{Cross-talk 2.0 studies. Laser fiber, 64-hole aperture, along with the internal structure of the MCP-PMT, including external voltage divider.}
\label{fig:crosstalk_2}
\end{figure}

Unlike cross-talk 1.0, which we could solve on a hardware level, in order to rectify this cross-talk 2.0 problem, we considered of a variety of software algorithms.
To make a detailed characterization of the charge-sharing effect, a special  
64-hole aperture was made that covers the glass of the MCP-PMTs, the holes being concentric with the pixels.
A laser fiber was inserted in one of the holes, as shown in Fig.~\ref{fig:crosstalk_2}, which also shows the schematics of the MCP-PMT.
The fiber was connected to the mTC laser calibration system, which contains a picosecond laser and filters for various attenuation levels. 
When the system was ready, a signal was sent to simultaneously trigger the laser and the readout electronics. 
For clarity, we call the pixel in which the laser is injected the {\it target} pixel, and the pulse generated in that channel {\it target} pulse.
We call the surrounding 8 pixels the {\it neighbor} pixels.

The goal was to make a transfer function that we could later apply to the real data. 
To the best extent of our knowledge, no one has yet developed such a function, as usually only a determination of a cross-talk at a certain level is being reported~\cite{DeFazio}. The effect is the most severe at higher gain.

\begin{figure}[!ht]
    \centering
    \includegraphics[width=.48\textwidth]{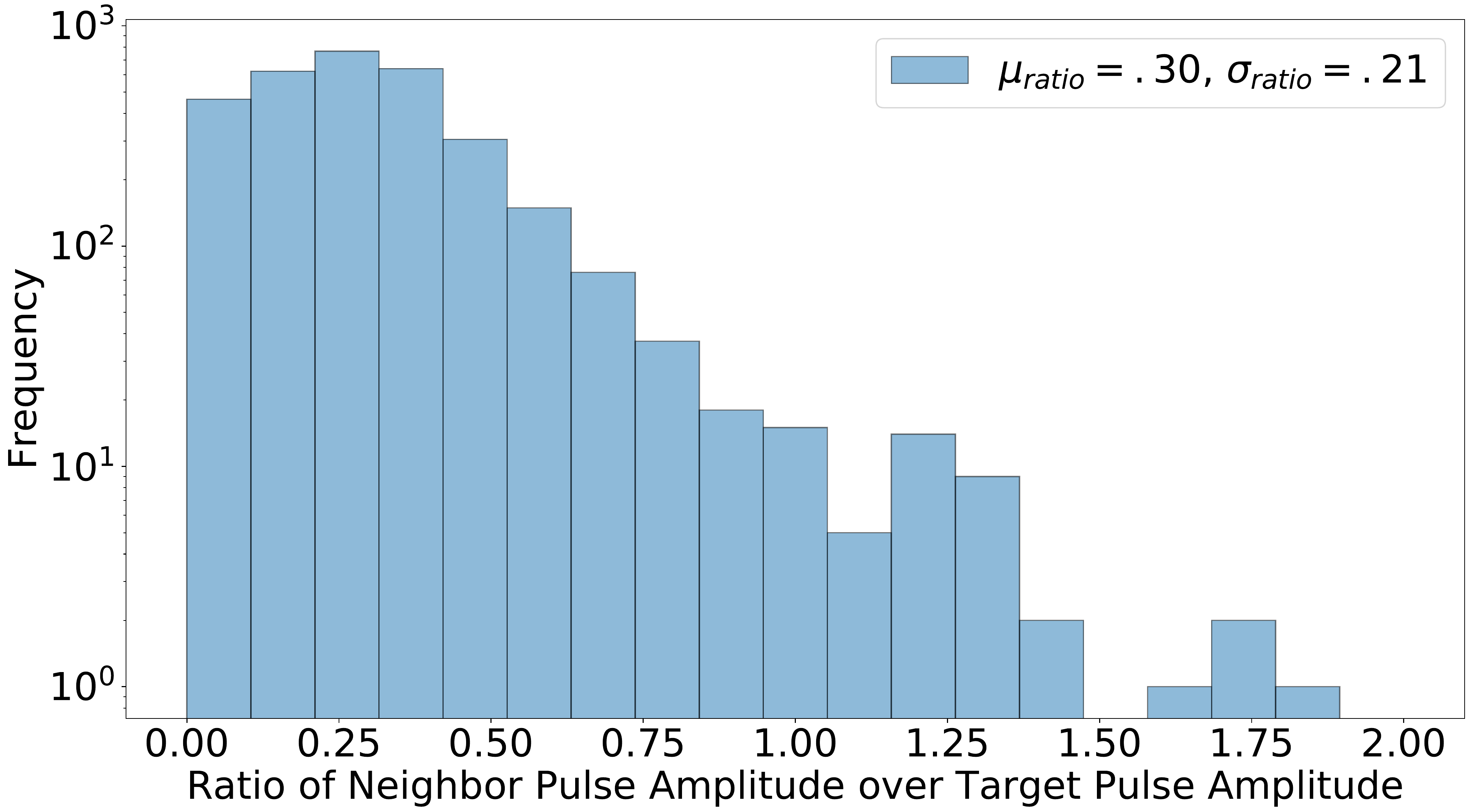}
    \caption{Relative Amplitude: Histogram showing the ratio of activated neighbor pulses to the amplitude of the target pulse. We see that this ratio averages at a significant 30\% with a standard deviation of 20\%.}
    \label{fig:Relative Amplitude}
\end{figure}

Analysis was done to ascertain the behaviour of the cross-talk quantitatively. By varying the MCP-PMT gain and laser attenuation and frequency, we were able to obtain many different sets of data which differed by frequency of hits and amplitudes. 
However, statistically, a pattern emerged in the cross-talk that was independent of varying these parameters. 
Fig.~\ref{fig:Relative Amplitude} is a histogram of the ratio of the amplitude of activated neighbor pulses to the amplitude of the target pulse for each event where both the target pulse and at least one neighbor surpassed 350 ADC counts. 
We see that the average value of the relative cross-talk amplitude is 30\% with a large spread. 
This effect is significant and cannot be ignored in any analysis which undertakes photon counting. 
These significant and inconsistent cross-talk effects were also observed by Vavra~\cite{vavra_talk}.

\begin{figure}[!ht]
    \centering
    \includegraphics[width=.48\textwidth]{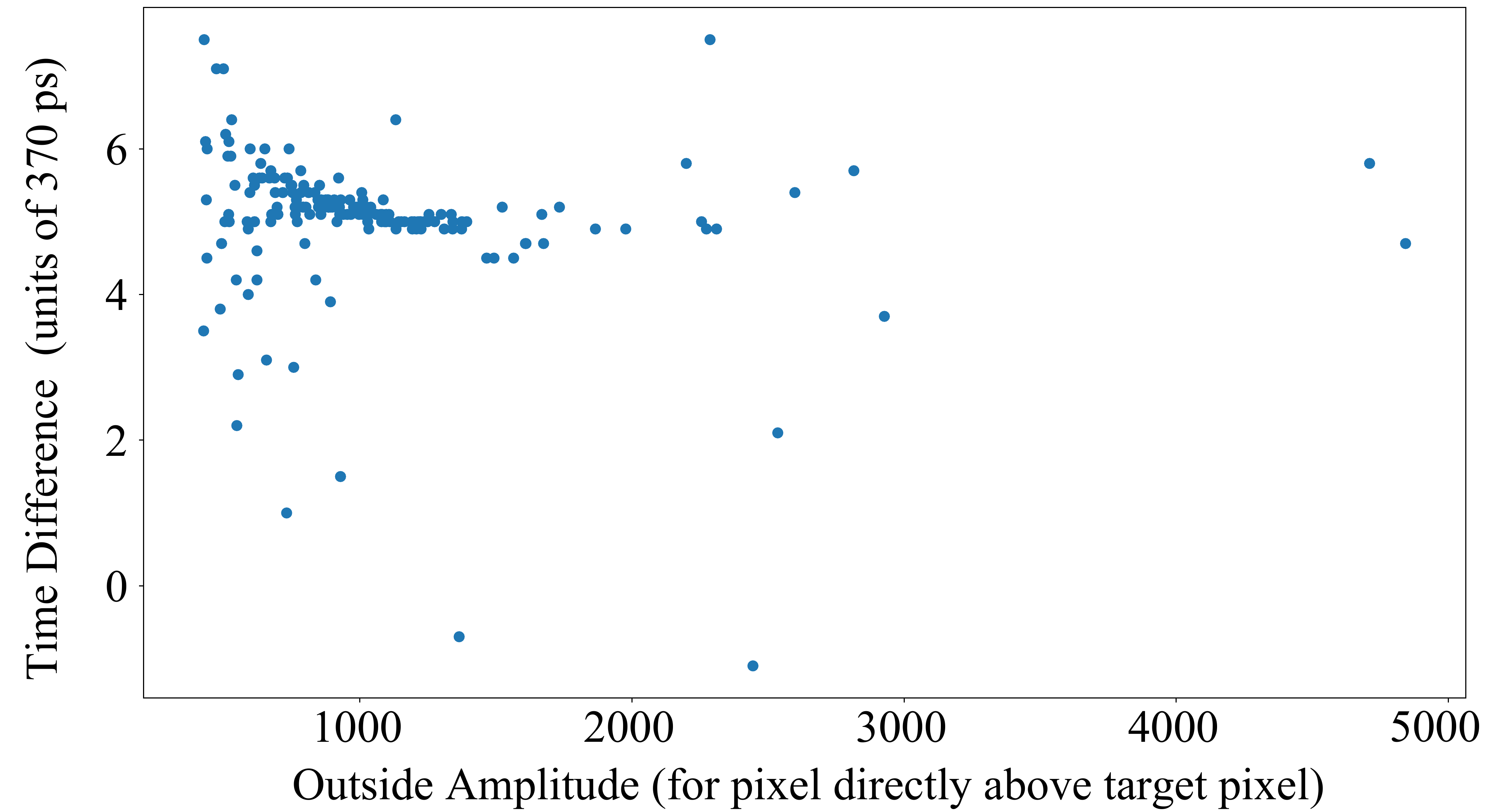}
    \caption{Timing Difference Between Neighbor Pulses And The Target Pulse: A scatter plot showing the time difference between the neighbor pulses and the target pulses plotted against the amplitude of that neighbor pulse. Any point that lands above zero on the y axis means that the neighbor pulse lagged in time. We see a general tendency for a small lag between the two pulse.  The tendency for greater and more consistent lag increases with the amplitude of the neighbor pulse. The average time difference independent of amplitudes is $1.7$~ns with a standard deviation of $1.1$~ns.}
    \label{fig:Time Difference}
\end{figure}

Next, we investigated the timing of the neighboring pulses compared to the target pulses in order to get a deeper understanding of the cross-talk. We were hopeful that a significant delay in the induced neighboring pulses could signify an easy way to eliminate them upon further analysis. Another reason for studying the timing differences was that it could either prove or disprove the hypothesis that these induced pulse were caused by capacitive coupling. 
If they were the result of capacitive energy transfer, they should behave as the derivative of the target pulse. 
Our results of this study are presented in the scatter plot of Fig~\ref{fig:Time Difference}. 
Here, we have the time difference between any activated neighbor pulse and the target pulse versus the amplitude of that neighbor pulse.  The fine time scale in the plot is on the order of 370~ps. 
We see that the timing difference between the target and neighbor pulses is quite small indeed, and that the neighbor pulses tend to lag by an average of 1.7~ns. This short delay supports the hypothesis that electron knockoff is the cause of the cross-talk, and eliminates the notion that the cross-talk is the result of capacitive coupling.
However, for fast timing purposes this delay is destructive.
For example, the mTC was built to be capable to detect neutron directionality
and order to achieve this we need to determine neutron recoil locations.
Our simulations show that the average time between consecutive recoils is $\sim 2$~ns,
thus, the phenomenon of cross-talk 2.0 completely inhibits our ability to determine the recoils by greatly diminishing our timing resolution.

To summarize, with respect to making a transfer function to correct for these timing delays,
there are two major problems. 
Firstly, the plot (Fig.~\ref{fig:Time Difference}) shows that the trend is inconsistent and that sometimes the neighboring pulses lead the target pulses. 
Secondly, the timing difference is too small to make confident cuts when the event requires fast timing. 
It is also notable that the larger the amplitude of the neighboring pulse, the greater the tendency for lag.  
In addition to using timing/amplitude corrections obtained from laser calibrations for this plot, we also made similar plots for different target pulse channels in the same region of the MCP-PMT, which showed similar results. 
These two steps helped to eliminate the potential for digitizer/routing delay variations in the electronics themselves to contaminate the results.

\begin{figure}[!ht]
    \centering
    \includegraphics[width=.48\textwidth]{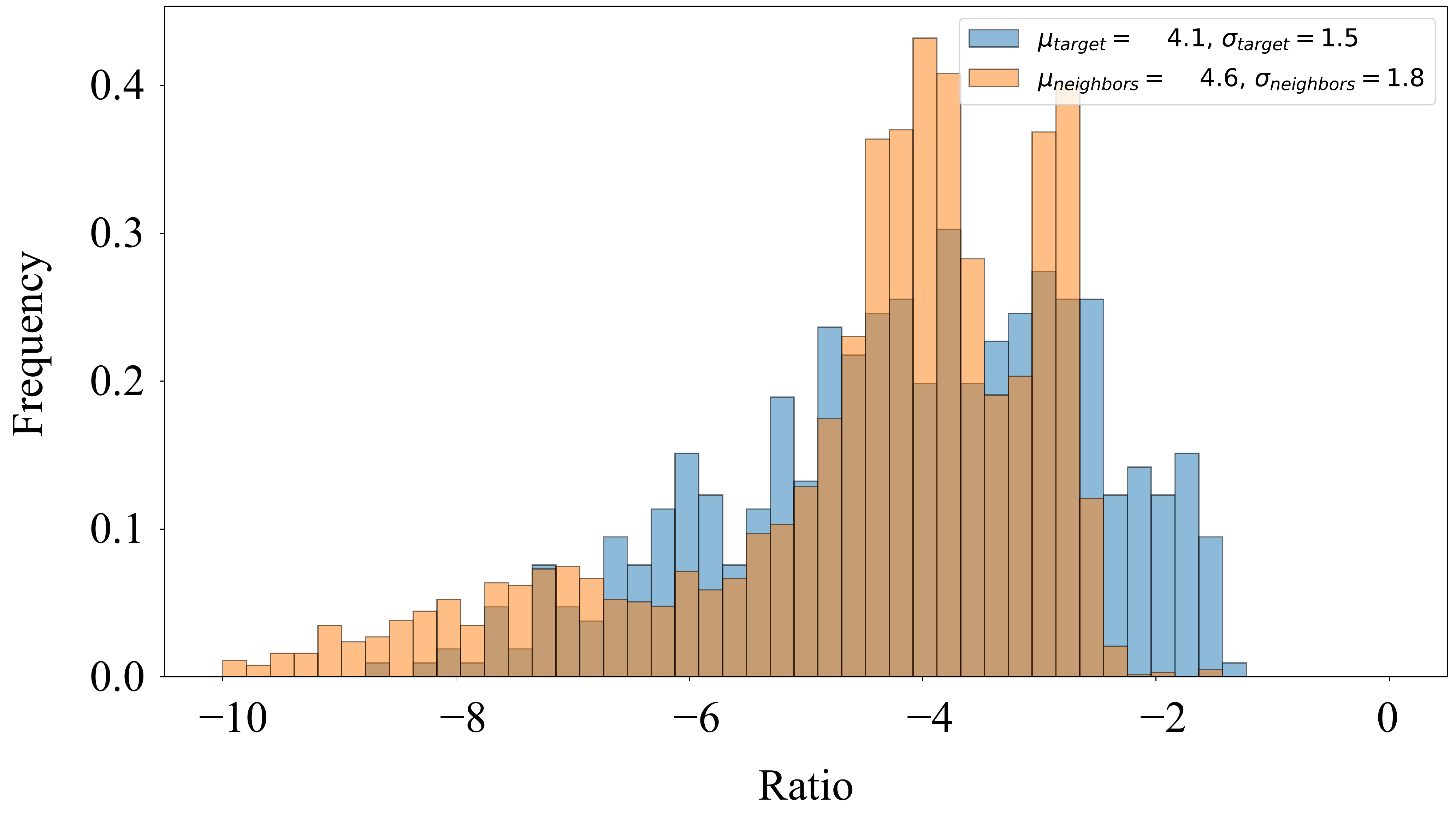}
    \caption{Ratio of Max to Min Amplitude of Individual Pulses: A Histogram comparing the distributions of the target pulse and activated neighboring pulse max/min ratios. The plot shows that the neighboring pulse ratios tend towards having less overshoot on average. However, this shows that there is much overlap in the two distributions and therefore it would be difficult to use this method to discriminate upon the pulse shape.}
    \label{fig:Pulse_Shape_1}
\end{figure}

Lastly, we looked to see if we could use difference in pulse shapes to eliminate the cross-talk. 
In Fig.~\ref{fig:Pulse_Shape_1} we show a histogram of the ratio of the maximum of each individual neighbor pulse to its minimum, and overlay it with the same ratio for each target pulse. 
Thereby, this is a study of how the neighbor pulse maximums relate to their overshoots (minimums) that is defined by $R=max/min$. 
The hope was to find a reliable statistical difference in pulse shape between the target and neighbor pulses.  
In Fig.~\ref{fig:Pulse_Shape_1}, the larger the negative value, the lesser the relative overshoot of the pulse. 
The results show a large amount of overlap between the two in terms of the ratios; however, the plot shows a slight bias for less/more overshoot for the neighbor/target pulses respectively. 
Again, as there is much statistical overlap, we do not see this as a reliable method to make cuts.

\begin{figure}[!ht]
    \centering
    \includegraphics[width=.49\textwidth]{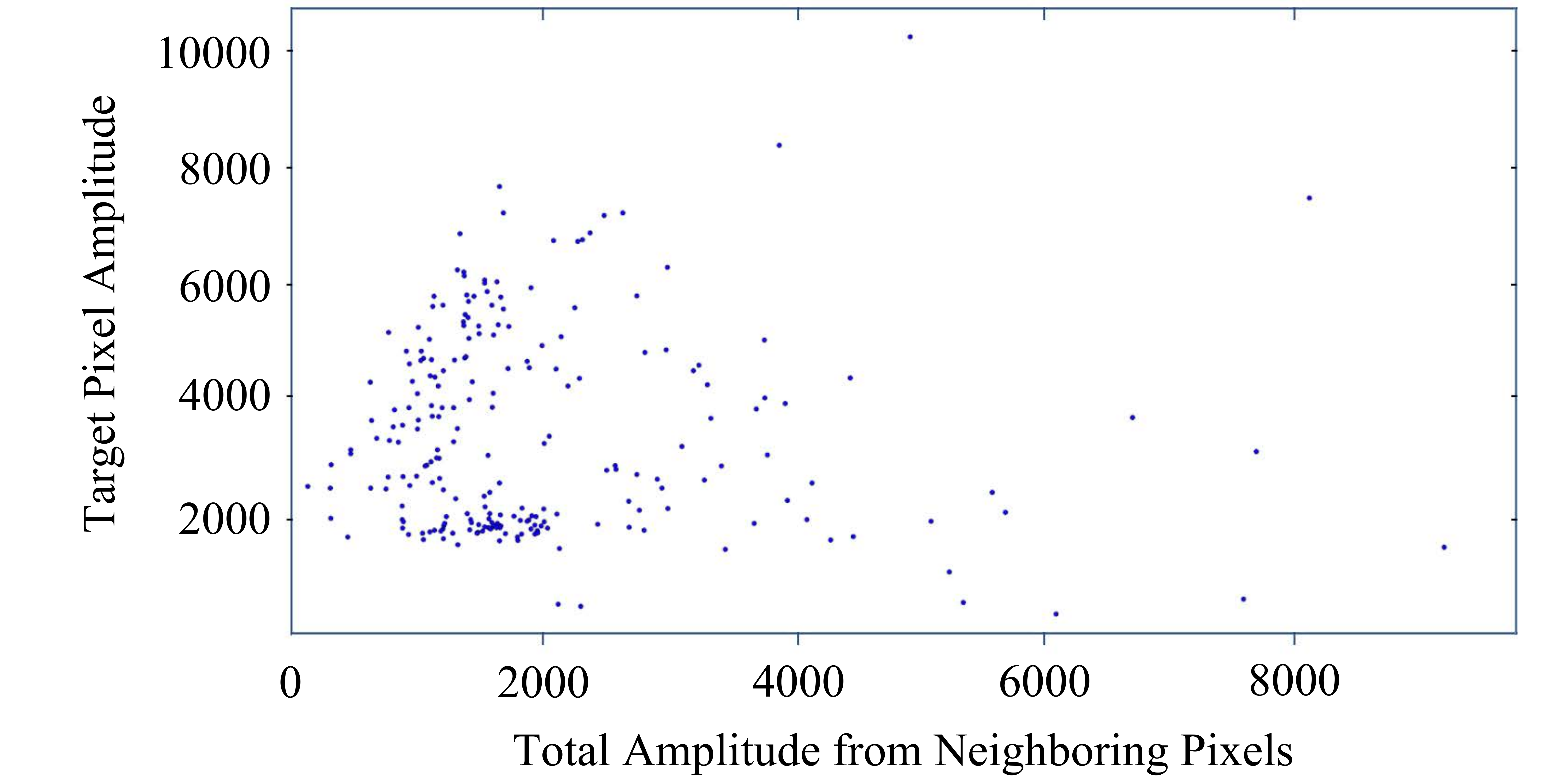}
    \caption{A scatter plot of the amplitude of the target pixel versus the total amplitudes of the surrounding pixels. The plot shows the wide varying response in the activity of the surrounding pixels to the charge placed in the target pixel. The ratio of the target pulse to the total amplitude of all 9 activated pixels (target plus neighbors) produced a value of $.575 \pm .270$.}
    \label{fig:totalamp}
\end{figure}

The ratio of the target pulse amplitude and the sum of all 8 neighboring pixels surrounding the target pixel, shown on the Fig.~\ref{fig:Pulse_Shape_1}, can be written as
\begin{equation}
    R = \frac{A_{\mathrm{target}}}{\sum \limits_{i=1}^8 A_i}
\end{equation}
The quoted value $R = .575 \pm .270$ is actually the ratio of the target amplitude to the sum of all 9 amplitudes of activated pixels (neighbor plus target), with the distribution shown in Fig.~\ref{fig:totalamp}.
We used this value to recalibrate our energy estimator, however, by doing so we introduced a large amount of uncertainty in our energy reconstruction.
Unfortunately, we were unable to develop a reliable transfer function for the timing. 

\section{Ion-feedback}
\begin{figure}[!t]
\centering
\includegraphics[width=.48\textwidth]{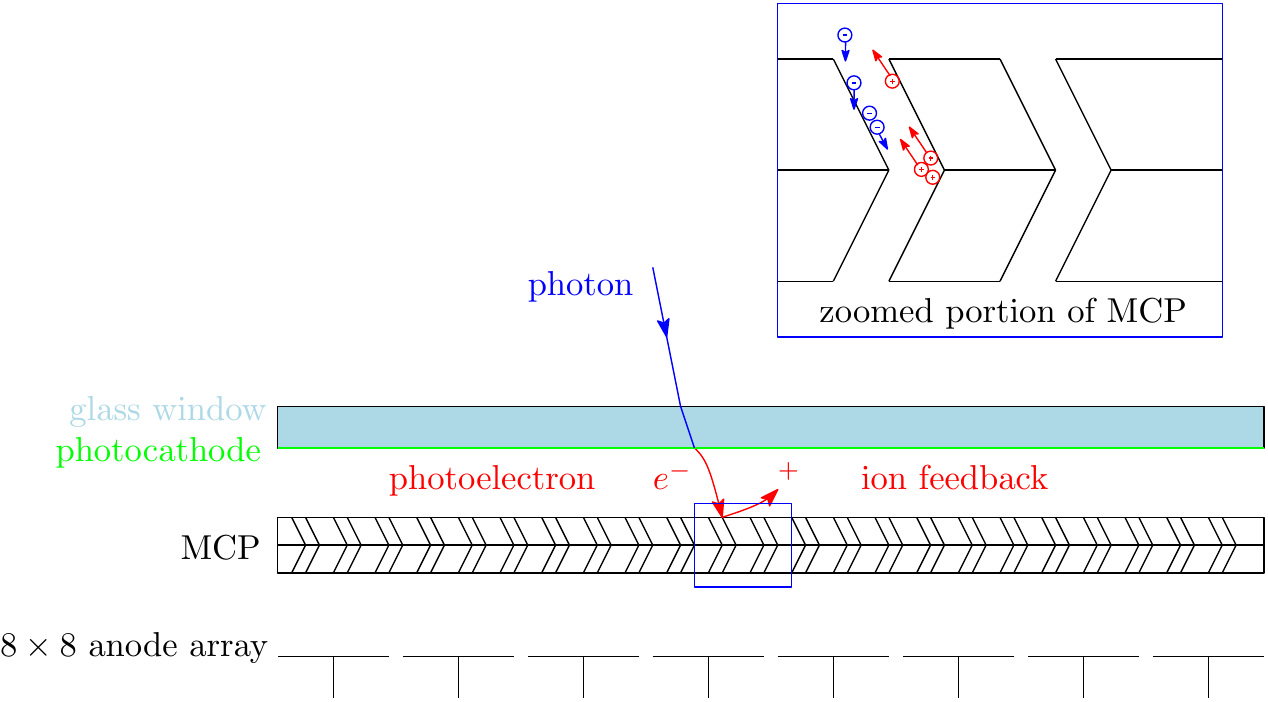}
\caption{MCP-PMT diagram showing ion feedback effect.}
\label{fig:ionfeedback}
\end{figure}

Another feature of MCP-PMTs is that there is always a chance that, along with electrons, there will be some ions produced in the porous MCP structure, as shown in Fig.~\ref{fig:ionfeedback}. Those ions, usually having energies in keV region, would propagate in the opposite direction to the electrons in the electric field with a chance of escaping the MCP, and flow back into it, causing a delayed signal to be read out on the multi-anode~\cite{vavra_talk,brandt_talk}. The delays can vary from a few nanoseconds to a few microseconds. It is relatively hard to solve this problem. The good thing about these delays is that they are not usually correlated among multiple MCP-PMTs or even among multiple groups of pixels within one MCP-PMT.
One potential solution to this problem is to use a different design for the MCP-PMTs, with a grid electrode installed between photocathode and microchannel plate~\cite{defazio_patent}.

\section{Conclusion}

Ultimately, cross-talk 1.0 
--- occurring between neighboring MCP-PMTs and affecting edge pixels ---
 was severely affecting the timing,
which is crucial for the compact detector like the mTC that utilizes first-photon arrival times. The issue was solved through the installation of copper frames.

Cross-talk 2.0 --- occurring between neighboring pixels within an individual MCP-PMT --- confuses energy calibrations and spatial mapping of events, especially when neighboring pixels are hit simultaneously.
Given our findings, MCP-PMTs are a non-ideal choice for low light measurements, particularly in the single-PE neighborhood.
Moving forward, SiPMs, LAPPDs, or other novel types of detectors may provide better alternatives.

\section{Acknowledgment}
We gratefully acknowledge the funding for the mTC project provided by National Geospatial-Intelligence Agency.

\section{References}\bibliography{references.bib}

\begin{thebibliography}{9}%
\makeatletter
\providecommand \@ifxundefined [1]{%
 \@ifx{#1\undefined}
}%
\providecommand \@ifnum [1]{%
 \ifnum #1\expandafter \@firstoftwo
 \else \expandafter \@secondoftwo
 \fi
}%
\providecommand \@ifx [1]{%
 \ifx #1\expandafter \@firstoftwo
 \else \expandafter \@secondoftwo
 \fi
}%
\providecommand \natexlab [1]{#1}%
\providecommand \enquote  [1]{``#1''}%
\providecommand \bibnamefont  [1]{#1}%
\providecommand \bibfnamefont [1]{#1}%
\providecommand \citenamefont [1]{#1}%
\providecommand \href@noop [0]{\@secondoftwo}%
\providecommand \href [0]{\begingroup \@sanitize@url \@href}%
\providecommand \@href[1]{\@@startlink{#1}\@@href}%
\providecommand \@@href[1]{\endgroup#1\@@endlink}%
\providecommand \@sanitize@url [0]{\catcode `\\12\catcode `\$12\catcode
  `\&12\catcode `\#12\catcode `\^12\catcode `\_12\catcode `\%12\relax}%
\providecommand \@@startlink[1]{}%
\providecommand \@@endlink[0]{}%
\providecommand \url  [0]{\begingroup\@sanitize@url \@url }%
\providecommand \@url [1]{\endgroup\@href {#1}{\urlprefix }}%
\providecommand \urlprefix  [0]{URL }%
\providecommand \Eprint [0]{\href }%
\providecommand \doibase [0]{http://dx.doi.org/}%
\providecommand \selectlanguage [0]{\@gobble}%
\providecommand \bibinfo  [0]{\@secondoftwo}%
\providecommand \bibfield  [0]{\@secondoftwo}%
\providecommand \translation [1]{[#1]}%
\providecommand \BibitemOpen [0]{}%
\providecommand \bibitemStop [0]{}%
\providecommand \bibitemNoStop [0]{.\EOS\space}%
\providecommand \EOS [0]{\spacefactor3000\relax}%
\providecommand \BibitemShut  [1]{\csname bibitem#1\endcsname}%
\let\auto@bib@innerbib\@empty
\bibitem [{\citenamefont {Li}\ \emph {et~al.}(2016)\citenamefont {Li} \emph
  {et~al.}}]{Li:2016yey}%
  \BibitemOpen
  \bibfield  {author} {\bibinfo {author} {\bibfnamefont {V.~A.}\ \bibnamefont
  {Li}} \emph {et~al.} (\bibinfo {collaboration} {mTC}),\ }\bibfield  {title}
  {\enquote {\bibinfo {title} {{Invited Article: miniTimeCube}},}\ }\href
  {\doibase 10.1063/1.4942243} {\bibfield  {journal} {\bibinfo  {journal} {Rev.
  Sci. Instrum.}\ }\textbf {\bibinfo {volume} {87}},\ \bibinfo {pages} {021301}
  (\bibinfo {year} {2016})},\ \Eprint {http://arxiv.org/abs/1602.01405}
  {arXiv:1602.01405 [physics.ins-det]} \BibitemShut {NoStop}%
\bibitem [{\citenamefont {Li}(4 05)}]{Li:2018dwm}%
  \BibitemOpen
  \bibfield  {author} {\bibinfo {author} {\bibfnamefont {V.}~\bibnamefont
  {Li}},\ }\emph {\bibinfo {title} {{miniTimeCube: Building The World's
  Smallest Neutrino Detector}}},\ \href {http://inspirehep.net/record/1666321}
  {Ph.D. thesis},\ \bibinfo  {school} {Hawaii U.} (\bibinfo {year}
  {2018-04-05})\BibitemShut {NoStop}%
\bibitem [{\citenamefont {Inami}\ \emph {et~al.}(2008)\citenamefont {Inami}
  \emph {et~al.}}]{INAMI2008247}%
  \BibitemOpen
  \bibfield  {author} {\bibinfo {author} {\bibfnamefont {K.}~\bibnamefont
  {Inami}} \emph {et~al.},\ }\bibfield  {title} {\enquote {\bibinfo {title}
  {{Cross-talk suppressed multi-anode MCP-PMT}},}\ }\href {\doibase
  https://doi.org/10.1016/j.nima.2008.04.014} {\bibfield  {journal} {\bibinfo
  {journal} {Nucl. Instrum. Meth.}\ }\textbf {\bibinfo {volume} {A592}},\
  \bibinfo {pages} {247 -- 253} (\bibinfo {year} {2008})}\BibitemShut {NoStop}%
\bibitem [{\citenamefont {Vavra}(2011)}]{Vavra:2010zz}%
  \BibitemOpen
  \bibfield  {author} {\bibinfo {author} {\bibfnamefont {J.}~\bibnamefont
  {Vavra}},\ }\bibfield  {title} {\enquote {\bibinfo {title} {{PID techniques:
  Alternatives to RICH Methods}},}\ }\href {\doibase
  10.1016/j.nima.2010.09.062} {\bibfield  {journal} {\bibinfo  {journal}
  {Nucl.Instrum.Meth.}\ }\textbf {\bibinfo {volume} {A639}},\ \bibinfo {pages}
  {193--201} (\bibinfo {year} {2011})}\BibitemShut {NoStop}%
\bibitem [{\citenamefont {Korpar}(2011)}]{Korpar11}%
  \BibitemOpen
  \bibfield  {author} {\bibinfo {author} {\bibfnamefont {S.}~\bibnamefont
  {Korpar}},\ }\bibfield  {title} {\enquote {\bibinfo {title} {{Electron
  backscattering in MCP}},}\ \ }(\bibinfo  {publisher} {{Photodetection:
  Principles, Performance and Limitations (CERN conference EDIT)}},\ \bibinfo
  {year} {2011})\BibitemShut {NoStop}%
\bibitem [{\citenamefont {DeFazio}()}]{DeFazio}%
  \BibitemOpen
  \bibfield  {author} {\bibinfo {author} {\bibfnamefont {J.}~\bibnamefont
  {DeFazio}},\ }\href@noop {} {\enquote {\bibinfo {title} {unpublished, private
  communication},}\ }\BibitemShut {NoStop}%
\bibitem [{\citenamefont {Vavra}(2016)}]{vavra_talk}%
  \BibitemOpen
  \bibfield  {author} {\bibinfo {author} {\bibfnamefont {J.}~\bibnamefont
  {Vavra}},\ }\href
  {https://indico.cern.ch/event/393078/contributions/2241767/} {\enquote
  {\bibinfo {title} {{Other PID techniques (9th International Workshop on Ring
  Imaging Cherenkov Detectors)}},}\ } (\bibinfo {year} {2016})\BibitemShut
  {NoStop}%
\bibitem [{\citenamefont {Brandt}\ \emph {et~al.}(2014)\citenamefont {Brandt}
  \emph {et~al.}}]{brandt_talk}%
  \BibitemOpen
  \bibfield  {author} {\bibinfo {author} {\bibfnamefont {A.}~\bibnamefont
  {Brandt}} \emph {et~al.},\ }\href@noop {} {\enquote {\bibinfo {title} {{Fast
  timing detector R\&D (MCP workshop)}},}\ } (\bibinfo {year}
  {2014})\BibitemShut {NoStop}%
\bibitem [{\citenamefont {DeFazio}(2016)}]{defazio_patent}%
  \BibitemOpen
  \bibfield  {author} {\bibinfo {author} {\bibfnamefont {J.}~\bibnamefont
  {DeFazio}},\ }\href {https://patents.google.com/patent/US9425030B2/en}
  {\enquote {\bibinfo {title} {Electrostatic suppression of ion feedback in a
  microchannel plate photomultiplier},}\ } (\bibinfo {year} {2016}),\ \bibinfo
  {note} {{US patent 9.425,030 B2}}\BibitemShut {NoStop}%
\end{thebibliography}%

\end{document}